\documentclass[twocolumn,pra,aps,showpacs]{revtex4}

\usepackage{psfrag,graphicx,overpic}

\usepackage{dcolumn}
\usepackage{amsmath,amssymb}
\usepackage{bm}
\usepackage{pstricks}

 \newcommand{\ket}[1]{|#1\rangle}
 \newcommand{\bra}[1]{\langle #1|}

\begin{document}


\title[Impurity Induced VBS in Cr8Ni]
{VBS State Induced by Impurity Frustration in Cr8Ni}

\author{
J. Almeida$^{\star}$, M.A. Martin-Delgado$^{\star}$ and G.
Sierra$^{\ast}$
 }
\affiliation{ $^{\star}$Departamento de F\'{\i}sica Te\'orica I,
Universidad Complutense. 28040 Madrid, Spain.
\\
$^{\ast}$Instituto de F\'{\i}sica Te\'orica, C.S.I.C.- U.A.M.,
Madrid, Spain. }

\begin{abstract}
We provide a physically meaningful picture of the nature of the
 ground state of the Cr8Ni compound in the regime where it is a spin singlet.
 According to this picture, the anisotropy 
of the $Ni$ atom in the $Cr$ ring induces a dimerization in the molecule that makes
the ground state to stabilize in a Valence Bond Solid phase of virtual spins.
We characterize rigorously
 this phase by means of a particular non-local order parameter denoted the generalized string order parameter. In the completely antiferromagnetic regime,
 the system becomes frustrated. We have performed a numerical 
real-time evolution study of the correlations
between the spin of the $Ni$ impurity and the rest of the spins 
in order to show the reaction
of the system under this frustration. We provide results that
 manifestly show the agreement with the Valence Bond Solid picture introduced.
\end{abstract}



\maketitle


\section{Introduction}
\label{sect_intro}
Recently, the physics of ring-shaped molecular magnets 
with antiferromagnetic interactions and an odd number
of interacting spin centers (e.g. paramagnetic ions)
has attracted a great deal of interest since they provide 
emblematic examples of systems where spin frustration effects due
to quantum magnetism play a major role. Moreover, they have been
sinthesized and studied experimentally 
\cite{Cr8Ni,heterometallic_Cr,ab_initio,yamamoto_08,yamamoto_03}.
Specifically, we shall concentrate on the anomalous magnetic properties in 
some heterometallic odd spin rings \cite{schnack_01,odd1,odd2,odd3}
namely, chromium rings. The system comprises eight chromium(III) ions 
with spin $S=\frac{3}{2}$ each, and one nickel(II) ion with spin $S=1$.
The magnetic properties of this first odd-member antiferromagnetic ring
has been investigated with electron paramagnetic resonance (EPR)
and its spin frustrated properties has been visualized by means of
a M\"{o}bius strip. In this paper we propose an alternative and
complementary picture of the ground state of  this Cr8Ni ring molecule
using valence bond states (VBS) \cite{aklt1,aklt2,fannes89,fannes91,fannes,korepin,rva,jaitisi,poilblanc08} of
virtual spins which are used to represent the spins $S=\frac{3}{2}, 1$
of the real constituent ions. We will show that the particular bond pattern
acquired by these VBS states is a consequence and manifestation of the
spin frustration in the odd Cr8Ni ring molecule.

Molecular nanomagnets are fascinating new magnetic materials
\cite{book2006,book2005,book2003}. They appear in a large
variety of compounds with many different properties. We shall
focus on antiferromagnetic compounds of bimetallic rings.
These molecules are ideal candidates to study the physics
of simple but nontrivial spin models like the AF Heisenberg 
interaction. The key point here is that, these molecules 
show very interesting finite-size quantum many-body effects which
are typically overlooked in other studies of the Heisenberg
model where the main focus is to achieve the thermodynamic
limit (number of spins going to infinity). In those studies,
the small finite-size effects are considered spurious effects
that vanish for larger and larger systems, which eventually
may show some sort of universality, if that is the case.
Quite on the contrary, the nice thing of these small molecules
is that we can vary its size and coupling constant strenghts
such that the finite-size effects become some real property
that can be addressed experimentally, theoretically 
and numerically. Some intersting examples
of these small quantum effects that we study in this paper
are level crossing, change in the nature of their the ground state
(e.g., from spin singlet to spin triplet or higher), existence of excited
states very close to the ground state, etc.

In Sect.\ref{Hamiltonian_description} we introduce a Heisenberg
Hamiltonian to describe the interactions between the two types
of ions in the bimetallic compound Cr8Ni, where the Ni ion plays
the role of an impurity within an homogeneous chain of $Cr$ ions with the
shape of a ring molecule. In fig.~\ref{spectrum} we present
the energy spectrum of this Hamiltonian obtained numerically with
an appropriate Lanczos technique.
In Sect.\ref{VBS_picture}, we first provide the VBS-picture
for the Cr8Ni ring molecule based on a strong coupling limit
in the Ni impurity coupling. This is the origin of the spin
frustration in the system. In order to support this VBS-picture,
we provide numerical results for a generalized string order 
parameter that is able to detect the type of virtual bond structure. 
In Sect.\ref{Frustration_Dynamics} we study the frustration effects in this ring-molecule and test the associated  VBS state picture by means of a numerical study of time-evolved correlation functions of spin-spin operators for different ions in the nano-molecule. Sect.\ref{sect_conclusions} is devoted to conclusions.


\section{Hamiltonian Description of the Cr8Ni Molecule Ring}
\label{Hamiltonian_description}

Magnetic molecules are emblematic instances of an ensemble of 
non-interacting quantum systems embedded in a 
condensed matter environment. 
The synthesis of molecular magnets has undergone rapid progress in recent years. 
Each of those quantum systems are identical molecular units 
that can contain as few as two and up to 
several dozens of paramagnetic ions (spins). 
In our case, they correspond to  Cr8Ni ring molecules.
This molecule is one of the many relevant molecules
containing transition-metal ions whose spins are so
strongly-exchange coupled that when the temperature
is low enough, their behaviour is like single-domain
particles with a certain total spin \cite{book2006,book2005,book2003}.

Macroscopically, these materials appear as crystals or
powders. Nonetheless, their intermolecular 
magnetic interactions are utterly negligible when compared to their 
intramolecular interactions.
Thus, measurements of their 
magnetic properties reflect mainly ensemble properties of single molecules. 
There are two major advantages in the research on these molecule aggregates. 
Firstly, the outstanding degree of accuracy by 
which their magnetic dynamics can usually be modeled.
Secondly,  the opportunity to chemically engineer molecules 
possessing desired physical properties \cite{book2006,book2005,book2003}.
In our case, the interest relies on the study of the Heisenberg model
in situations that are usually discarded when studying that model
in infinite one-, two-, and three-dimensional systems.
Our studies will reinforce the idea that such spin arrays 
yield qualitatively new physics caused by the finite size of the system.

The Cr8Ni compound belongs to a wider family of molecular rings. In molecules
with a small number of ions, there exist big differences in their physics depending on
each particular compound. A first major difference is related to the number of its 
ions:  odd or even. In this regard, the implications are mainly twofold: first,
having a difference of one ion in the same family of molecules,  can cause 
the molecule to have a neat magnetic moment or not,  therefore changing its magnetic properties 
drastically. Second, and interestingly enough,  a molecular ring with completeley
antiferromagnetic interactions between nearest
neighbors ions can be a candidate to present quantum spin frustration 
properties if the number of atoms is odd, while 
this effect will not generally be present for even member rings in a given family compound.

As it happens,  in the majority of these molecules 
the localized single-particle magnetic moments of the ions
couple antiferromagnetically. Then, their spectrum is described 
rather well by the Heisenberg model with very few parameters
because of the high symmetry of the molecular configurations.
These coupling parameters correspond to
isotropic nearest neighbor interaction sometimes augmented by anisotropy terms.

The Cr8Ni compound is one of the first antiferromagnetic odd member rings 
which has been artificially sinthesized. 
The results of its magnetic properties are interpreted within the framework 
of a spin Hamiltonian approach and they nicely fit the 
pattern of the energy levels obtained by inelastic neutron spectroscopy.
There exists also reports on its magnetic and spin 
frustration effects \cite{odd1}. In view of these properties, it has been
proposed \cite{Cr8Ni} that the behavior of this molecule can be 
properly explained with a nearest neighbors Heisenberg model where only two different
microscopic couplings play a role: one is the coupling that parametrizes the strength
of the interaction existing between the $Ni$ and the neighbor pair of $Cr$. The other
one is the coupling that takes into account the interaction between the $Cr$-$Cr$ pairs, which
can be considered the same for each pair.  The easy-axis anisotropy term is reported
to be very weak how as to play any role.
Therefore, the Hamiltonian that we shall study has the following form:
\begin{equation}
\begin{split}
H & = J\sum_{i=1}^7 \textbf{S}_{\rm Cr}(i)\cdot \textbf{S}_{\rm Cr}(i+1) \\
  & + J'\left[ \textbf{S}_{\rm Ni}\cdot \textbf{S}_{\rm Cr}(1) + 
\textbf{S}_{\rm Ni}\cdot \textbf{S}_{\rm Cr}(8)\right],
\end{split}
\label{Hamiltonian}
\end{equation}  
where, for convenience, the $Cr$ atoms have been numbered from 1 to 8, being these
latter the two neighbors of the $Ni$ atom. Notice that, since the spin of the niquel is equal to
$\textbf{S}_{\rm Ni}=1$ and the spin of the $Cr$ atoms is $\textbf{S}_{\rm Cr}=3/2$, the total
spin of the molecule must be integer. 

\begin{figure}
\includegraphics[width=9cm]{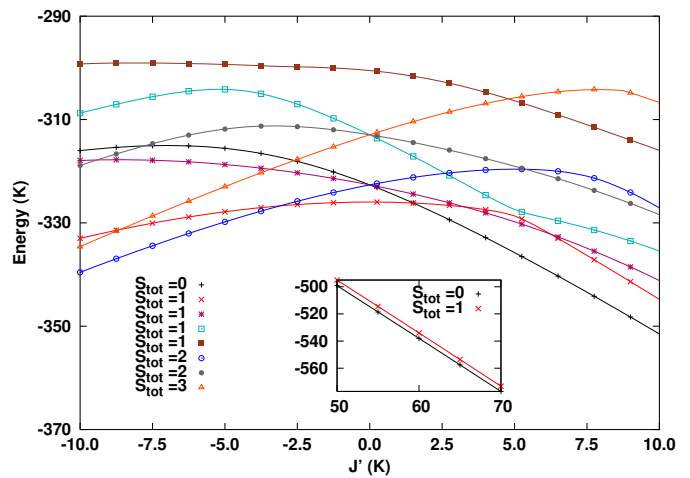}
\caption
{
Energy spectrum of some of the lowest lying states in each sector with well defined total spin. 
In the coupling region shown, the spectrum is highly dependent on the values of $J$ and $J'$.
In particular, in the antiferromagnetic region $J'>0$ K the ground state is a triplet
for $J'<1.5$ K and a singlet elsewhere. \emph{Inset:} for higher values of $J'$ the 
ground state is a singlet and the first excited state is a triplet very close in energy.
}
\label{spectrum}
\end{figure}


In fig.~\ref{spectrum} we have plotted some lowest lying energy levels of this Hamiltonian.
In Sect.~\ref{Frustration_Dynamics} we shall explain how these numerical results have been obtained
with a multitarget Lanczos method.  
It can be observed that, in some regime of the coupling constants $J'$ and $J$,
the energy levels are highly braided and, as a consecuence, the
ground state has different value of the total spin depending on the exact value
of these couplings. However, as can be seen from the inset of this figure, for large
values of $J'$ the ground state is always a spin singlet with a triplet state very close
in energy above it. In this work, we will restrict our study to the antiferromagnetic
($J'>0$) region and in particular to the region where the ground state is a singlet. From
fig.~\ref{spectrum} we see that this area corresponds to $J'>1.5$ K, while the region 
$0<J'<1.5$K is characterized by a ground state with total spin equal to one 
(for convenience the computations in fig.~\ref{spectrum} have been done with a fixed value
of $J=16$ K).

The interest in the domain where the ground state is a singlet comes not only from the
fact that it spans the most extension in the antiferromagnetic area, but also because
the physics of the real Cr8Ni seems to be in agreement with a regime close to
$J=16$ K and $J'=70$ K, with a non-magnetic ground state.
 Therefore, the interest of our study relies on the fact that it can provide
new insights into the physics of a not so well-known state of matter but
 with a well defined connection with experiments in real compounds.
   
\section{Impurity induced VBS picture}
\label{VBS_picture}

A valence bond solid is a particular quantum many-body state that can be understood as
follows: given a system of \emph{real} particles with total spin $S$, we
 can split each one of them into $2S$ \emph{virtual} particles of
 spin $S=1/2$. In order to recover the original spins, we enforce these 
\emph{virtual} particles to couple (i.e., symmetrize) among themselves 
in order to give the original
spin $S$ particles. 
To create now a wavefunction with total spin equal to zero, we make
singlets (i.e., antisymmetrize) out of every pair of \emph{virtual} particles. 

We will denote each of these singlet pairs between \emph{virtual} particles as a bond.
There are a lot of different possible ways to fix the bonds between all the 
\emph{virtual} particles and, in general, the total wave-function may have contributions from
all these configurations. There exist however some physical situations in which
only some particular bond configurations, out of the whole possible set, take part in
the wave-function: some systems have a major contribution coming only from one particular
bond arrangement. These systems are commonly dubbed bond crystals. It may also happen that there
exist not one but some few bonds configurations whose weights are dominant in the total
wave-function. In this case the system is called a resonating valence bond solid (RVBS).
There exist yet another kind of more disordered states, denoted $(m,n)$-VBS, which
 we shall see that describe properly the ground state of the Cr8Ni in the singlet region.
A general $(m,n)$-VBS state is built by  forming bonds only
 between \emph{virtual} spins belonging to neighbor \emph{real} particles, with
 the numbers $m,n$ satisfying $m+n=2S$ and $S$ being the spin of the \emph{real} particle.

\begin{figure}
\includegraphics[width=6cm]{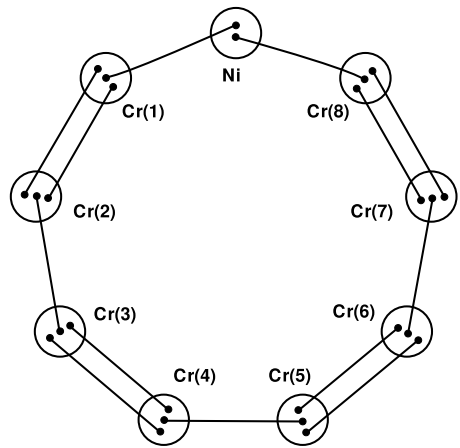}
\caption
{
Valence Bond Solid picture of the Cr8Ni ring molecule. The bonds established between the
two virtual spins in the $Ni$ and the two virtual spins in the neighbor $Cr$ atoms forces
a dimerized pattern in the rest of the chain. This VBS configuration becomes the
dominant one in the strongly coupled impurity limit $J'\gg J$.
}
\label{VBSCr8Ni}
\end{figure}

These states are usually translationally invariant (with the
$(m,n)$-VBS notation this means that $m=n$) in those systems where the
Hamiltonian possess this symmetry. There also exist dimerized
$(m,n)$-VBS states that have been shown to appear in systems where the
full translational symmetry of the Hamiltonian has been partially
broken such that still exist an enlarged unit cell. Typically, this
effect can be obtained by introducing an external dimerization
coupling constant in the Hamiltonian that still preserves some
periodicity.

Our main result in this section is that there is another mechanism to provide
such dimerized $(m,n)$-VBS states in the Cr8Ni ring molecule
and whose success precisely resides in the existence of an impurity
within an homogeneous system.
To understand this mechanism in the particular case of the Cr8Ni, we resort
 to the strong coupling limit where
the antiferromagnetic interaction between the $Ni$ and its neighbors is much larger that
the interaction among $Cr$ pairs. As shown in fig.~\ref{VBSCr8Ni}, the two
 \emph{virtual} spins comprising the $Ni$
will be likely to form bonds with the \emph{virtual} particles in the neighbor chromiums to
satisfy their antiferromagnetic constraints. The rest of the \emph{virtual} spins left
will then tend to form similar bonds with their neighbor partners, giving as a result a dimerized 
non translationally invariant VBS. We would like to stress the fact that the validity of
this picture is rooted in the existence of the $Ni$ impurity. In fact, the physics
of a homogeneous system of $Cr$ atoms is closer to a gapless critical phase rather than
to such a gapped state. 

General $(m,n)$-VBS states belong to a class of spin liquids which are
known to possess an special hidden order that can be identified by a
particular non-local order parameter called the string order parameter
(SOP) \cite{sop1,sop2,oshikawa92,sop_ladders,xiang08,top}.  This order
parameter has proved itself extremely successful in the task of
characterizing diverse kinds of such states, both in the pure one
dimensional cases and also in less trivial systems such as ladders
\cite{sop_ladders}, \cite{top}.  We shall see that this parameter
also allows us to characterize the Cr8Ni ground state.  The definition of
the generalized string order parameter is as follows
\cite{oshikawa92}:

\begin{equation}
O_{\textrm{str}}(\theta)=\lim_{\vert j-i\vert \rightarrow\infty} \langle
S^z_{i}\textrm{exp}(i\theta\sum_{k=i}^{j-1}S_k^z)S_{j}^z\rangle.
\label{SOPdefinition}
\end{equation}

\begin{figure}[h]
\includegraphics[width=9 cm]{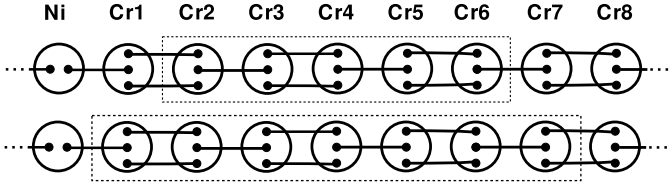}
\caption{Representation of two ways to measure the SOP in the Cr8Ni ring
choosing  adjacent starting sites.
In the diagram at the top we break two bonds at the left end and one at the right. This
corresponds to a $(2,1)$-VBS. In the diagram at the bottom we have the same 
physical configuration but now we are breaking one bond at the left and two at the right end,
which characterizes a $(1,2)$-VBS. Notice that both measures have been 
taken with an odd number of sites, otherwise the SOP is zero.}  
\label{odd_evenVBS}
\end{figure}

The limit $\vert j-i\vert$ going to infinite must be understood as 
comprising a region large enough so as 
to neglect finite-size efects. In our case the system itself is finite,
 however we will see that the results are still conclusive despite
some corrections that have to be considered due to this fact.

We hereby summarize the most relevant properties of the SOP for our purposes:\\
\noindent
i/ The SOP is symmetric with respect $\theta=\pi$.\\
ii/ The imaginary part of the SOP vanishes as we consider larger systems.\\
iii/ Due to the fact that the definition of the SOP makes use of antisymmetric operators under
spin flip, the number of sites considered in the measure must be odd in order to have an even
number of these operators. Otherwise the SOP is zero.\\
iv/ Given a generic $(m,n)$-VBS state, the number of zeros of this operator in
the interval $\theta \in[0,2\pi)$ coincides with the number $m$.\\
v/ Two measures of the SOP begining in 
adjacent sites will differ in the order of the numbers $m$ and $n$. That is, if
one measure gives a $(m,n)$-VBS state, the other will be a $(n,m)$-VBS (see fig.~\ref{odd_evenVBS}).\\

\begin{figure}[ht]
\begin{overpic}[width=9.0cm]{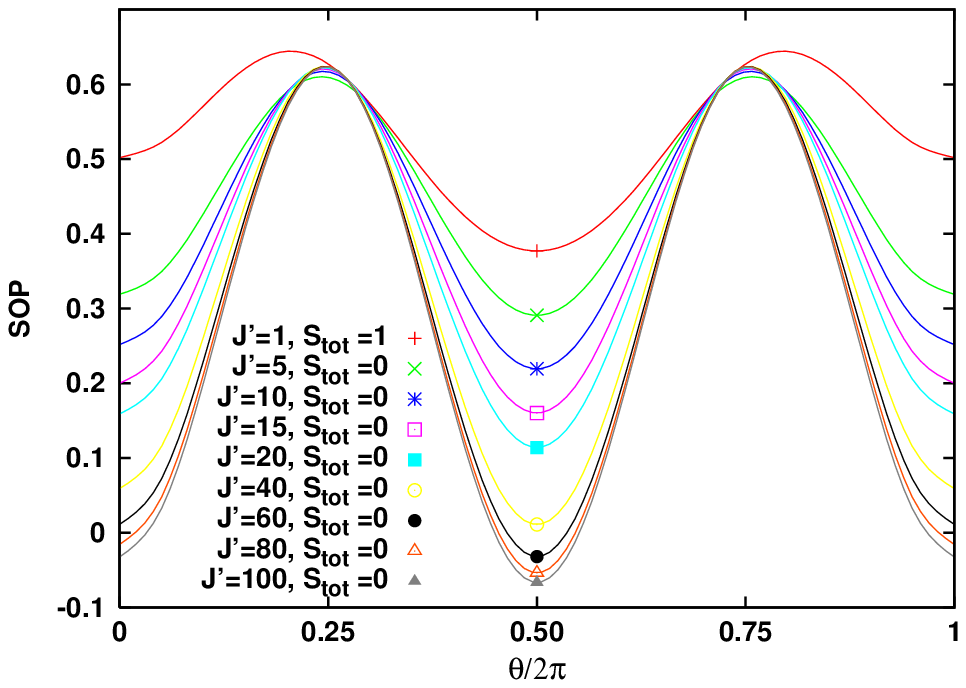}
\put(0,65){a)}
\end{overpic}
\begin{overpic}[width=9.0cm]{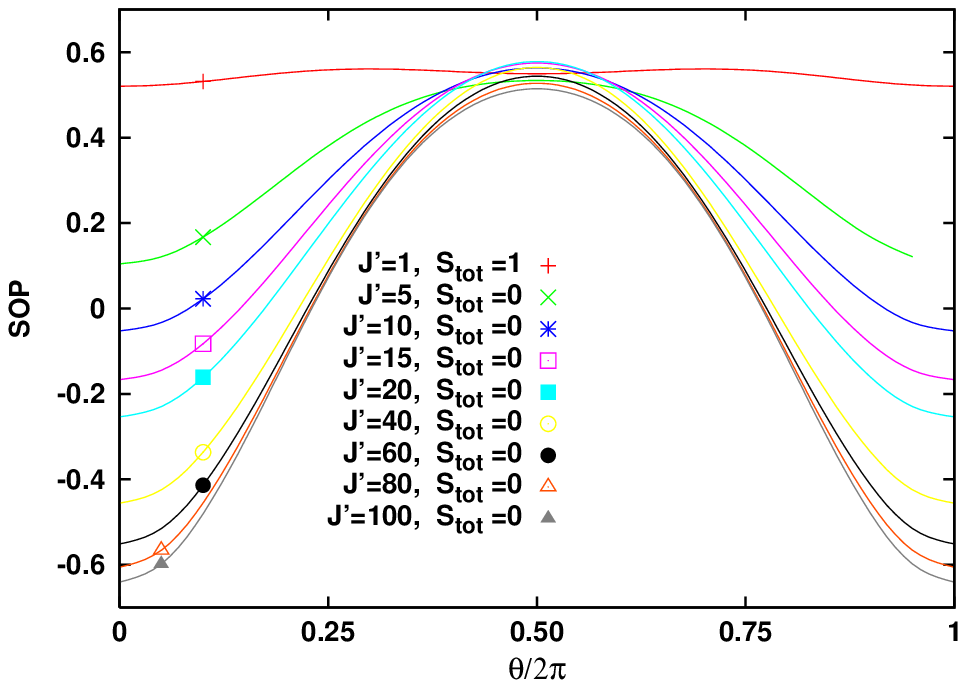}
\put(0,65){b)}
\end{overpic}
\caption
{ a) SOP computed between the $Cr_2$ and the $Cr_6$ atoms. b) SOP
computed between the $Cr_1$ and the $Cr_7$ atoms. We have fixed
$J=16$. Considering the interval $\theta\in [0,2\pi)$ and looking at the
curves with high values of J' we see that the number of zeros in a) is
two, corresponding to a $(2,1)$-VBS. In b) the situation is more
confusing due to stronger finite size effects that shifts the curves
towards negative values at $\theta=0$ and $\theta=\pi$. With this
correction in mind the graph strongly suggests the existence of a
$(1,2)$-VBS.  }
\label{SOPfig}
\end{figure}

In order to show the existence of a VBS-picture of the Cr8Ni
molecule as described in fig.\ref{VBSCr8Ni},
we have computed  the generalized string
order parameter in the ground state of the molecule for different
 ratios of the coupling constants $J$ and $J'$. In virtue of property iv/ above we
can characterize the VBS state counting the number of zeros of this operator. Moreover,
we have some freedom at choosing the starting and end sites to measure the SOP (i.e, 
the numbers $i$ and $j$ in definition \eqref{SOPdefinition}) and we will see 
that the results are consistent with property v/.
 
In fig.~\ref{SOPfig} we have plotted the SOP computed in two different
blocks of the ring and adjacent starting sites
 (these blocks are represented in fig.~\ref{odd_evenVBS}). Remarkably, even in this small molecule,
 the imaginary part of the SOP vanishes in all our computations.\\
 In fig.~\ref{SOPfig}a the curves with high values of $J'$ have two local minima in the interval
 $\theta\in[0,2\pi)$ whose value is compatible with zero considering that the system
 is finite and the block used to measure the SOP comprises only some few spins. This result
is consistent with a $(2,1)$-VBS. The shape of the SOP is also typical of these states,
with two noticeable maxima placed approximately at $\theta=\pi/4$ and $\theta=3\pi/4$.
Accordingly, the computations shown in fig.~\ref{SOPfig}b shall be consistent 
with a $(1,2)$-VBS as 
explained above. We see that again the shape of the SOP for the strong coupling curves
shares the main features of these states, that is, one substantial maximum placed precisely
at $\theta=\pi/2$. There is however a major difference at $\theta=0$ and $\theta=\pi$, where 
the SOP does not vanish but has a significant negative value. At these points the
exponentials in the definition of the SOP do not play any role and hence it is a usual
spin-spin correlator between the two $Cr$ atoms at the end of the block. Since 
our ring is finite and closed it is quite natural to think that the 
$Cr_1$ and $Cr_7$ spins are highly correlated via 
the spin frustration in the $Ni$ impurity. Therefore, such a finite value at $\theta=0$
 only shows the compromise between the bulk $(1,2)$-VBS picture of the whole system 
and the local physics of those spins.

In summary, the string order parameter reveals that the ground state of 
the Cr8Ni in the strong impurity coupling limit is consistent with a
 impurity-mediated mechanism of spin frustration, the result of which
is the VBS-pattern shown  in fig.~\ref{VBSCr8Ni}. Since
we are dealing with a finite system the possibilities to drastically change the nature
of the ground state are limited, that is, the possibility of a quantum phase transition
is excluded and only a crossing with another energy level can produce this effect.
 Therefore, the intermediate $J'/J$ regime can
be considered as some deformation of the strong coupling limit. As we decrease $J'$ and
the $S_{tot}=1$ state crosses the $S_{tot}=0$, the VBS picture breaks down and the measures
of the SOP are not meaningful in the sense that the properties of this 
operator in such a state are not well defined.

\section{Frustration and Dynamics in Cr8Ni}
\label{Frustration_Dynamics}

As we have already mentioned, the Cr8Ni ring molecule is frustrated in the sense that
the minimum energy of the system can not be obtained minimizing separatedly each
of the two body terms of the Hamiltonian \eqref{Hamiltonian}. Another way to see it
is by resorting to the classical limit where each spin is pictured as a classical vector. 
Once we set the 
value of one of those classical spins, then we can fix one by one the rest of the spins in order to minimize the
local interactions, but in the end there will be one spin for which the local interaction
 with both of its neighbors can not be minimized at the same time.

 Typically, systems where frustration exists come along with a  
 rich and very often not so well-known physics. Roughly speaking, we can say
 that frustration in general increases the complexity of those systems, both in the
physics they exhibit as well as  in the way to approach them. In particular, there 
is not a well defined way to measure the amount and localization 
 of frustration. An attempt to quantify these effects in Cr8Ni can be
done attending to the structural changes of the gound state as we vary the 
couplings. That is, by inspection of the way in which spins in the ring couple to form the final
 state. This procedure has a connection with experimental techniques where the Lande factors
 of the ring can be measured. However this procedure is not suitable to study a rotationally 
invariant singlet ground state where the spin is zero. In this section we will study the
behavior of the Cr8Ni molecule by means of computing the time evolution of some important
spin correlators: the spin autocorrelation of the impurity $Ni$ atom with itself and
the spin correlation between the $Ni$ atoms and each $Cr$ along the ring. 
These correlators correspond to the vacuum expectation value 
of the time-evolved spin operators  $\textbf{S}_{\textrm{Ni}}(t)$ and $\textbf{S}_{\textrm{Cr}_i}(t)$
projected onto the spin operator of the $Ni$ impurity at $t=0$, $\textbf{S}_{\textrm{Ni}}(0)$.
This is a way to dynamically probe \cite{balseiro_87,balseiro_88,dagotto_93,karen1,karen2} the spin structure
in the ground state $\ket{\psi_0}$ of the ring molecule. In fact, we shall consider the square modulus of
those correlators and interpret them as time-evolution probabilities.
That is, we shall consider the following correlators in order to construct a figure of merit:
\begin{equation}
C_{\textrm{Ni}}(t):=
\bra{\psi_0}\textbf{S}_{\textrm{Ni}}(t)\cdot
\textbf{S}_{\textrm{Ni}}(0)\ket{\psi_0},
\label{Nicorrelator}
\end{equation}
and 
\begin{equation}
C_{\textrm{Cr}_i}(t):=
\bra{\psi_0}\textbf{S}_{\textrm{Cr}_i}(t)\cdot
\textbf{S}_{\textrm{Ni}}(0)\ket{\psi_0},
\label{Crcorrelator}
\end{equation}
Notice that in a rotationally invariant singlet ground state $\ket{\psi_0}$  the correlations in the $x,y$ and
$z$ axis have the same value and thus,  the expresions above can be written as:
\begin{equation}
C_{Ni}(t)= 3 \bra{\psi_0}S^z_{\textrm{Ni}}(t)
S^z_{\textrm{Ni}}(0)
\ket{\psi_0},
\end{equation}
\begin{equation}
C_{Cr_i}(t)= 3 \bra{\psi_0}S^z_{\textrm{Cr}_i}(t)
S^z_{\textrm{Ni}}(0)
\ket{\psi_0},
\end{equation}
Since the proportionality factor does not provide any additional information we will discard it from now on and will consider the bare $z$-axis projection correlators. The time dependency of the operators is given by the usual Heisenberg 
picture:
\begin{equation}
O(t)=e^{iHt}O(0)e^{-iHt}.
\end{equation}
The idea behind this figure of merit to measure the dynamical correlations
between spins is similar to the static correlator used to measure spin correlations
in space separated sites $i$ and $j$ of the ring $\bra{\psi_0}\textbf{S}_{\textrm{Cr}_i}(i)\cdot
\textbf{S}_{\textrm{Ni}}(j)\ket{\psi_0}$. This static correlator measure spatial correlations,
while our purpose is to measure time-evolved correlations which will probe not only the ground
state physics but also the excited states physics.


 We next explain briefly the numerical method used to evaluate these correlators.
 After that we shall show and discuss the results.

\subsection{Numerical Method}

\begin{table}
\begin{tabular}{|c|r|r|}
\hline
$\textbf{S}_{tot},\pm S^z_{tot}$ & Dimension & Dimension \\
\hline
0 & 1000 & 23548 \\ \hline
1 & 2764 & 22548 \\ \hline
2 & 3905 & 19784 \\ \hline
3 & 4256 & 15879 \\ \hline
4 & 3900 & 11623 \\ \hline
5 & 3095 & 7723 \\ \hline 
6 & 2150 & 4628 \\ \hline 
7 & 1308 & 2478 \\ \hline 
8 & 692 & 1170 \\ \hline 
9 & 314 & 478 \\ \hline 
10 & 119 & 164 \\ \hline 
11 & 36 & 45 \\ \hline 
12 & 8 & 9 \\ \hline 
13 & 1 & 1 \\ \hline 
Total: & 196608 & 196608 \\
\hline
\end{tabular}
\caption
{
Dimension of each subspace with well defined quantum numbers out of the
 total Hilbert space of a Cr8Ni ring. The second
column corresponds to the subspaces with well defined total spin. The third one are
the sectors with well defined value of the $z$-axis projection of the total spin. In this
last case for each value in the first column we must consider the positive and negative
cases.
}
\label{sectordims}
\end{table}

The Hamiltonian \eqref{Hamiltonian} is SU(2) rotational invariant. That is, it commutes both with the
total spin and the $z$-axis projection of the total spin. In table \ref{sectordims} we show the
dimensions of the subspaces corresponding to the conserved quantum
numbers of these operators. These sizes are in the limit to perform
exact diagonalization but lie however within the
domain reachable for a Lanczos method. 


To do our computations we will make use of an adapted version of the 
Lanczos algorithm specific to compute real-time dynamics \cite{prelovsek_94,prelovsek_95,review_finite_T}. 
In this framework
it can be shown that the correlators written above can be expressed as:
\begin{equation}
\begin{split}
C_i(t)=\sum_{j=0}^M
\bra{\psi_0}S^z_{Cr(i)}(t)
S^z_{Ni}(0)\ket{\tilde{\psi_j}}\times \\
\times\bra{\tilde{\psi_j}}
S^z_{Ni}(0)
\ket{\psi_0}
e^{-i(\tilde{\epsilon_j}-E_0)t},
\end{split}
\end{equation}
where $M$ stands for the dimension of the Krylov space $\mathcal{K}(H,q_0,M)$ 
such that $\mathcal{K}(H,q_0,M)=\mathcal{K}(H,q_0,M+1)$, with $q_0:=S^z_{Ni}\ket{\psi_0}$.
That is, $M$ is the
dimension of the largest invariant subspace generated by succesive
 aplications of the Hamiltonian 
$H$ upon the seed vector $q_0=S^z_{Ni}\ket{\psi_0}$. The vectors $\ket{\tilde{\psi_j}}$ are
the aproximated eigenvectors computed in this Krylov subspace, $\tilde{\epsilon}_j$ are 
the energies of these eigenvectors and $E_0$ stands for the energy of the ground state.

The number $M$ is typically
much lower than the total dimension of the Hilbert space, but still high to numerically
 compute a complete basis of the Krylov subspace. Therefore, the approximation in this method
resides in the fact that we will substitute the dimension $M$ with a lower number of vectors
that still serve as a complete basis for these correlators.

In order to obtain the most accurate results and representations of the eigenvectors 
of the Hamiltonian $\ket{\tilde{\psi_j}}$ we have not used the same Krylov space
to compute them all. Instead we have performed a Lanczos iteration to find the
ground state. After that, the Lanczos iteration is restarted with the previously found
 eigenvectors projected out of the subspace to find the next excited state. And so on
and so forth,
 until we have computed the desired number of eigenvectors. In particular, to compute
the correlators described before we have used 400 eigenvectors of the Hamiltonian. As regards of
the tolerance in the eigenvalues, we have set it to $10^{-14}$ allowing a maximum dimension
of each Krylov space of 350 vectors. Were we in an exact situation, these vectors should be normalized to
one and be orthogonal among themselves. Let us call $V$ the matrix whose columns
are these eigenvectors. We have checked that we obtain the following accuracy, 
\begin{equation}
\|^tVV-\mathbf{1}\|\sim 10^{-4},
\end{equation}

which can be considered a low value for such a large number of eigenvectors. 
Moreover, as another check of the accuracy of the eigenvectors we have computed the total spin
of each one of them and we have obtained integral values up to precisions of $10^{-6}$ 
in the vast majority of them. As for the invariance of the Hamiltonian under reflection
with respect the $Ni$ atom, we have checked that symmetric one and two body correlators
evaluated on every eigenvector are the same up to the fourth or fifth decimal digit.

\begin{table}[h]
\begin{tabular}{|r|r|r|r|}
\hline
J'& $C_{Ni}(0)$& Error (\%) \\
\hline
2 &   0.66666    & 0.00002 \\ \hline
10&   0.66666    & 0.0009 \\ \hline
20&   0.66653    & 0.02 \\ \hline
30&   0.66512    & 0.2  \\ \hline
40&   0.65660    & 2    \\ \hline
50&   0.57469    &14    \\ \hline
\end{tabular}
\caption
{
Relative error between the Ni self-correlation at $t$=0 using \eqref{Nicorrelator}
 with M=400 eigenstates and the value 
$\bra{\psi_0}S^z_{\textrm{Ni}}S^z_{\textrm{Ni}}\ket{\psi_0}$ in the
ground state, which is equal to $2/3$ .
}
\label{c_error}
\end{table}

\begin{figure}[h]
\begin{overpic}[width=6.0cm]{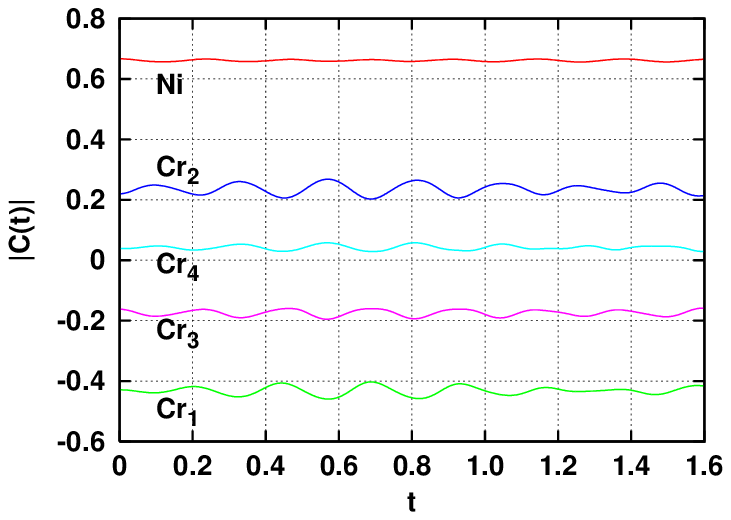}
\put(-5,65){a)}
\end{overpic}
\begin{overpic}[width=6.0cm]{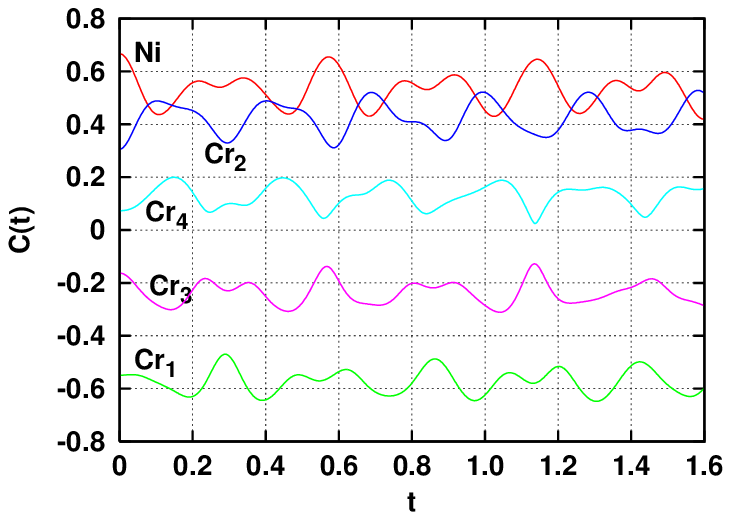}
\put(-5,65){b)}
\end{overpic}
\begin{overpic}[width=6.0cm]{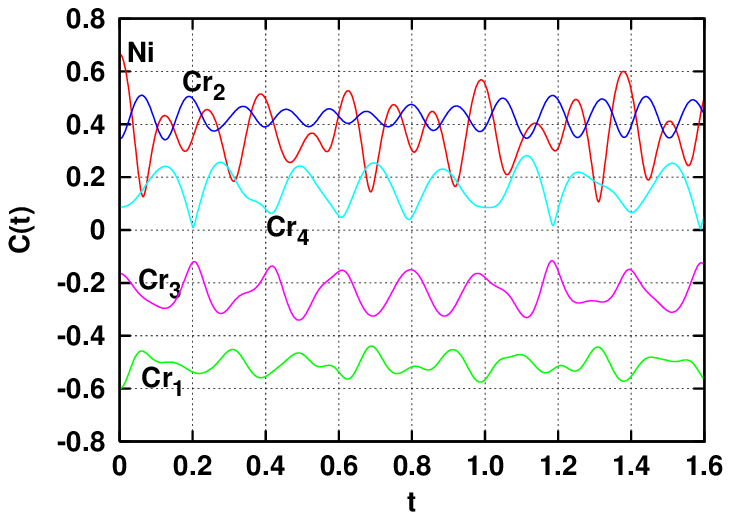}
\put(-5,65){c)}
\end{overpic}
\caption
{
Modulus of the spin correlation of the impurity $Ni$ atom with 
itself and with the rest of the $Cr$ atoms. The
graphs have been done with a value $J=16$ K and: a) $J'=2$ K, b) $J'=10$ K, c) $J'=20$ K. For convention
the sign of the correlators has been chosen to coincide with the sign of the correlation
at t=0, which is a real number.
}
\label{dynamics1}
\end{figure}

In order to check how complete is our set of eigenvectors, we have compared the value
 at $t=0$ computed using \eqref{Nicorrelator} and \eqref{Crcorrelator} with 
the values obtained measuring the correlators in the ground state 
without the projectors in between. We have observed that the agreement
 is excelent for low values of $J'$ while it goes worse for higher values
 of the coupling constant. Table \ref{c_error}
 shows these values and the relative error.

These accuracy checks confirm that our numerical results are good enough so as to
interpret them on physical grounds with respect to the spin frustracion effects
described in previous sections.

\subsection{Results}
\begin{figure}[h]
\begin{overpic}[width=6.0cm]{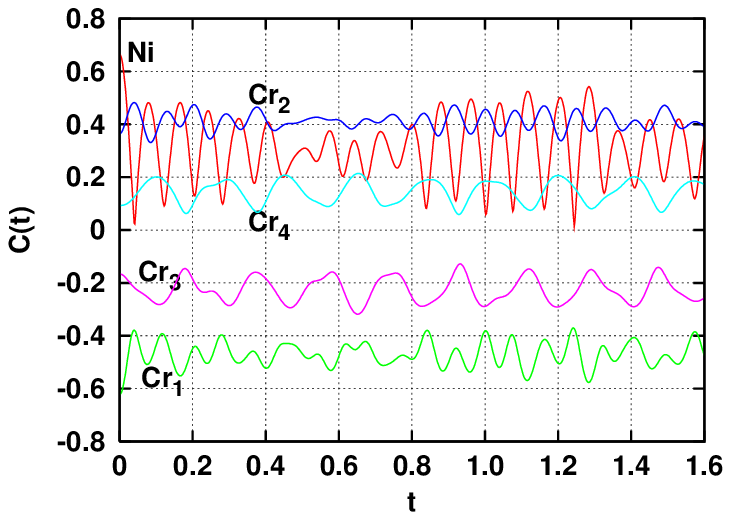}
\put(-5,65){a)}
\end{overpic}
\begin{overpic}[width=6.0cm]{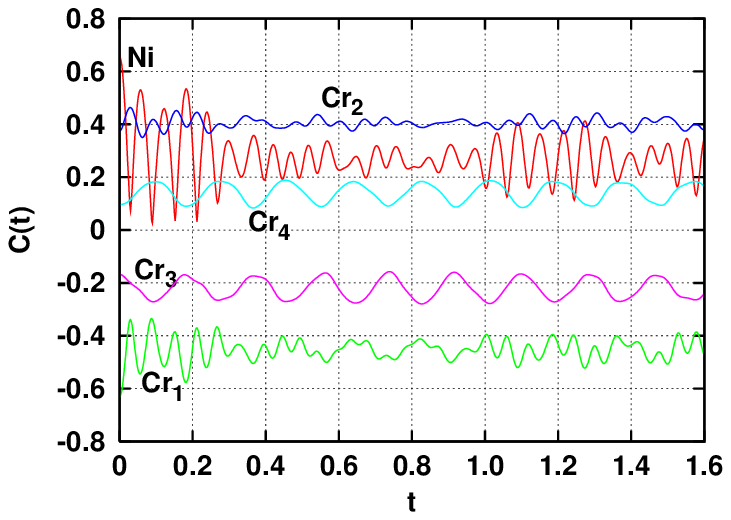}
\put(-5,65){b)}
\end{overpic}
\begin{overpic}[width=6.0cm]{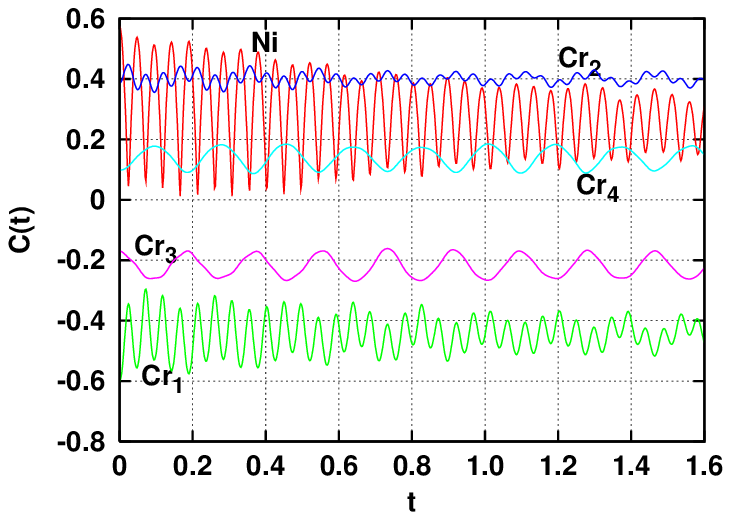}
\put(-5,65){c)}
\end{overpic}
\caption
{
Modulus of the spin correlation of the impurity $Ni$ atom with 
itself and with the rest of the $Cr$ atoms. The
graphs have been done with a value $J=16$ K and: a) $J'=30$ K, b) $J'=40$ K, c) $J'=50$ K. For convention
the sign of the correlators has been chosen to coincide with the sign of the correlation
at $t=0$, which is a real number.
}
\label{dynamics2}
\end{figure}

In the previous sections we have proposed and checked with the proper order parameters
a static picture of the ground state of the Cr8Ni ring. In the limit when the $Ni$ is 
weakly coupled to the $Cr$ bulk ($J'/J\ll 1$) the system is accurately described by
an isolated $Ni$ atom and an open $Cr8$ chain. On the other hand,  when the impurity
is strongly coupled ($J'/J \gg 1$) the ring posess a dominant contribution
in the form of a $(2,1)$-VBS ground state 
with some local correlations around the $Ni$ atom due to the finite size of the sample.
From the point of view of frustration, in this limit the impurity acquires strong 
antiferromagnetic compromises with both of its neighbors that can not satisfy simultaneously.
Frustration is known to impose complex constraints that can destabilize, deform and even 
produce new states of matter. On the other hand, the dynamics of each spin of the
system is highly influenced by these constraints. In the following paragraphs
we will see that the self correlation of the $Ni$ impurity and the rest of
correlators with the $Cr$ atoms allows us to naturally establish a relation
 with the amount of frustration.

In figs.~\ref{dynamics1} and \ref{dynamics2} we have plotted the
correlators (\ref{Nicorrelator}) and (\ref{Crcorrelator}) for a fixed
value of the constant $J=16$ K and different values of $J'$. The Cr8Ni
ring is invariant under reflection respect the impurity site and
therefore we will only provide the correlators with the $Ni$ itself
and the $Cr$ atoms numbered from 1 to 4 (with the notation of
fig.~\ref{VBSCr8Ni}), the correlations with the $Cr$ atoms numbered
from 5 to 8 are the same as their symmetric counterparts. We want to
provide these magnitudes with the meaning of a time-evolved
probability and hence we will consider only their modulus. It is worth
noticing that the time correlators mentioned before, at $t=0$ are real
numbers whereas for arbitrary values of $t$ they are complex
numbers. For convention, in the graphs where we plot the modulus of
these correlators we will provide them with the same sign of their
real value at $t=0$ to make explicit the ferromagnetic or
antiferromagnetic nature that they possess in the static $t=0$ ground
state.

In fact, we can observe in these graphs that the fingerprint of an
antiferromagnetic order is present in the initial $t=0$ ground state
and shows up in the alternation of the signs of the
correlators. Notice also a signal of frustration in the fact that this
alternation fails in the $Cr_4$ and its symmetric counterpart $Cr_5$
(due to the reflection invariance of the ring their value is equal
with the same sign), where the correlations reveal that both spins are
oriented in the same direction respect the spin of the
$Ni$. Remarkably this ferromagnetic defect is a consecuence only of
the reflection invariance of the Hamiltonian.

From these graphs we can also infer that the average correlation of each spin with
the impurity is little sensitive to the strength of the coupling constant $J'$, although
the amplitude of the deviations with respect to this average value increases with it.

\begin{figure}
  \includegraphics[width=8cm]{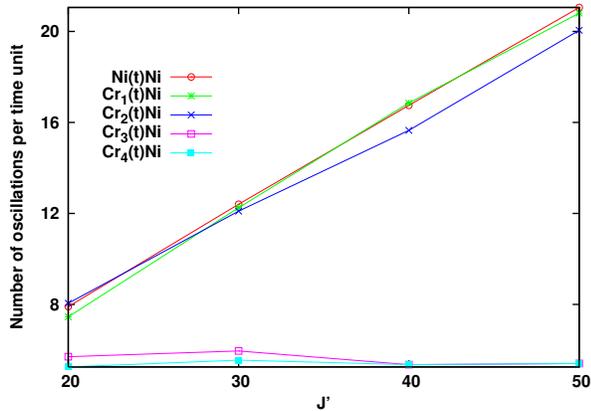}
\caption
{ 
The number of oscillations has been obtained counting the number
of minima in a wide time interval and dividing by the total time.
The Ni, $Cr_1$ and $Cr_2$ have increasing frecuencies with $J'$ which
corresponds to more frustrated regimes. In particular for the Ni spin
the relation of these variables is linear up to a high precision. In the 
case of the $Cr_3$ and $Cr_4$ spins the frecuency is hardly affected 
by $J'$.
}
\label{fig_oscillations}
\end{figure}

The most important observation is that the dynamics of the
correlators exhibit a non trivial sort of periodicity. That is, from
the shape of the curves it seems that there exists many modulating
components but a dominant pattern of oscillations is
apparent. Moreover, the frecuency of this pattern clearly increases
with increasing values of the impurity coupling $J'$, i.e. as we move
towards more frustrated regimes, but not in the same way for all the
spins. We have captured in fig.\ref{fig_oscillations} the frecuency of the
 dominant oscillatory pattern of each correlation. The graph highlights
two different tendencies of the correlations depending on the considered spins:
the frecuency of the $Ni$ self correlation as well as
 the correlations of the $Ni$ spin with the $Cr_1$ and 
$Cr_2$ spins increase with $J'$. Moreover in the case of the $Ni$ self correlation 
the relation of these two variables is linear with a surprising accuracy. On the other hand
the $Cr_3$ and $Cr_4$ are less affected by the impurity spin and the frecuency
remains almost constant in the wide range considered.

\begin{figure}[h]
\includegraphics[width=6.0cm]{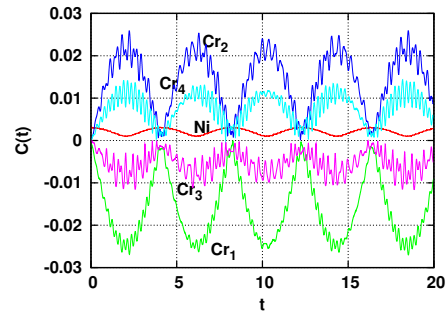}
\caption
{
Modulus of the spin correlation of the impurity $Ni$ atom with 
itself and with the rest of the $Cr$ atoms. The values of the coupling constants are $J=16$ K and $J'=1$ K
which corresponds to the region where the ground state is a triplet. For convention
the sign of the correlators has been chosen to coincide with the sign of the correlation
at $t=0$, which is a real number.
}
\label{dynamics3}
\end{figure}

In fig. \ref{dynamics3} we have plotted the same real time correlations with an election of
the coupling constants $J=16$ K and $J'=1$ K such that the ground state is a triplet.
In this case the ground state is not rotationally invariant  and the $z$-axis projection of the correlators is not
proportional to the scalar correlators \eqref{Nicorrelator} and \eqref{Crcorrelator} . For our purposes however, these magnitudes suffice to realize the different nature of both the singlet and triplet ground states: first of all is that not one but two dominant patterns of oscillation are well distinguishable in the triplet regime. Secondly and also a major difference is that the scale in this regime is some orders of magnitude smaller than the singlet case. 

The existence of this oscillatory behavior seems natural in a frustrated system where there
does not exist a natural equilibrium position for each spin or where the resulting
equilibrium configuration may result unstable. In a system composed by classical spins
these oscillations can be interpreted as the necessary movements of each spin to satisfy
the frustrated interactions, becoming faster as we blur the concrete equilibrium positions
 with the frustrating interactions.
 
The results in fig. \ref{fig_oscillations} point towards a regime where the frustration
introduced by the $Ni$ impurity has strong local dynamical effects in the nearest and
next nearest $Cr$ neighbors while the rest of the spins perceive the impurity screened
 by this closer shell of atoms and therefore their dynamics is little affected by it.
These results also show that the correlators proposed to study the frustration of the
system indeed have the behavior expected for a suitable estimator in order 
to measure the intuitive idea we have about the amount of frustration in a certain system.

\section{Conclusions}
\label{sect_conclusions}


  In recent years, considerable efforts have been devoted
to synthesizing and investigating magnetic systems 
of nano scale dimension that comprise a controllable number 
of transition metal ions. Highly symmetrical clusters of almost planar 
ring shape are among such topical molecular 
nanomagnets. In particular, the bimetallic ring molecule Cr8Ni
is the first antiferromagnetic ring with an odd number of spins.
Thus, it is a remarkable quantum system to test fundamental magnetic properties, 
and in particular the spin frustration effects.
 
In this work, we have studied the Cr8Ni frustrated ring in the regime where
the ground state is a singlet. That is, with a fixed value $J=16$ K this region corresponds 
to $J'>1.5$ K. In this regard, the experimental characterization of a Cr8Ni 
molecule places the real strengths present in the real system close to $J=16$ K and $J'=70$ K, well
within the singlet region.

 As we let the interaction strength of the $Ni$ impurity to be 
 stronger than that between the $Cr$ atoms, the ring stabilizes in a ground state with
the quantum properties of a dimerized VBS. The picture that explains this behavior in terms of
the possible bonds between neighbor particles comes clear from fig.~\ref{VBSCr8Ni}. 
Such a VBS state constitutes an example of a spin liquid with an intrinsic order that
can be measured by means of some particular non-local order parameters. In fig.\ref{SOPfig}
 we show the computations of this order parameter on the ground state and 
its behavior supports neatly the VBS picture. In this regard, some finite size effects
can be observed in the order parameter that reveal a competition between the physics
in the bulk of the ring and the strong effects, possibly mediated by the system frustration,
that ocurrs in the vicinity of the $Ni$ atom.

In the second section of this paper we have studied the role of the frustration in such
a VBS state by means of computing the real time evolution of the spin correlators between
the atoms in the ring. In particular, we have found that the amount of
frustration can be related to the frecuency in the oscillatory behavior 
of this correlators. This relation can be naturally established from the observation
that the oscillations in the system become faster as we move to the more frustrated regime
$J'\gg J$. Such an oscillatory behavior is natural in a system where 
no natural equilibrium is allowed due to the 
frustration. However, the spin correlators reveal that the atoms that are
most affected by this frustration are the $Ni$ impurity itself and those $Cr$ atoms that are closer
to it, that is Cr$_1$ and Cr$_2$, while
the effect of the impurity strenght seems to be less influent in the Cr$_3$ and Cr$_4$ atoms.

We believe that the methods and numerical techniques used in this work are versatile enough
and can be extended to a variety of other nanomolecular magnetic compounds.

\vskip4ex

\noindent {\em Acknowledgements}: Part of the computations of this
work were performed with the High Capacity Computational Cluster for
Physics of UCM (HC3PHYS UCM), funded in part by UCM and in part with
FEDER funds.  We acknowledge financial support from DGS grants under
contract FIS2006-04885 and the ESF Science Programme INSTANS
2005-2010.


\end{document}